%Paper: hep-th/9208021
%From: NEPOMECHIE%PHYVAX@umiami.ir.miami.edu
%Date: 07 Aug 1992 18:43:06 -0400 (EDT)

% Macro is already included; just TEX twice to get the references right
%
%

\newif\ifproofmode			% true => wide right margin, eqn.
\proofmodefalse				% labels shown; date in headline

\newif\ifforwardreference		% true => allow reference to an
\forwardreferencetrue			% equation that appears later

\newif\ifeqchapternumbers		% true => equations labeled as e.g.
\eqchapternumbersfalse			% (3.7) instead of (7)

\newif\ifsectionnumbers			% true => sections will be numbered.
\sectionnumberstrue			% Must be true if the next are true.

\newif\ifeqsectionnumbers		% true => equations labeled with
\eqsectionnumbersfalse			% section numbers: (4.7) or (3.4.7)
					% The last form occurs if BOTH
					% eqchapter and eqsection numbering
					% is enabled.

\newif\ifchaptersectionnumbers     	% true => section numbers as 3.4
\chaptersectionnumberstrue		%         (chapter.section)

\newif\ifcontinuoussectionnumbers	% true => don't reset numbering in eqs.
\continuoussectionnumbersfalse	% in each section (if
					% \eqsectionnumberstrue)

\newif\ifcontinuousnumbers		% true => don't reset numbering of
\continuousnumbersfalse 		% equations in each chapter

\newif\iffigurechapternumbers		% true => figures labeled as e.g.
\figurechapternumbersfalse		% (3.7) instead of (7)

\newif\ifcontinuousfigurenumbers	% true => don't reset numbering of
\continuousfigurenumbersfalse		% figures in each chapter

\newif\ifcontinuousreferencenumbers     % Normally, all references are
\continuousreferencenumberstrue         % packaged together

\newif\ifparenequations			% Default is parentheses around
\parenequationstrue			% the equations referred to

\newif\ifstillreading			% To check for eof in input

\font\eqsixrm=CMR6			% given these odd names to avoid
			% possible conflict with a user.
\def\marginstyle{\eqsixrm}		% for proofmode labels

\newtoks\chapletter			% for appendices
\newcount\chapno			% chapter number
\newcount\sectno			% section number
\newcount\eqlabelno			% equation #'s
\newcount\figureno			% figure numbers
\newcount\referenceno			% footnote counter
\newcount\minutes			% time of day stamp
\newcount\hours				%   "

\newread\labelfile			% input name for labels
\newwrite\labelfileout			% output "    "    "
\newwrite\allcrossfile			% output "    "  cross references

\chapno=0
\sectno=0
\eqlabelno=0
\figureno=0

% A few definitions made here simply to guarantee upward compatibility.

\def\chapternumberstrue{\eqchapternumberstrue}

% The initialization procedure.
%
\def\initialeqmacro{
    \ifproofmode
        \headline{\tenrm \today\ --\ \timeofday\hfill
                         \jobname\ --- draft\hfill\folio}
        \hoffset=-1cm
        \immediate\openout\allcrossfile=zallcrossreferfile
    \fi
    \ifforwardreference
        \openin\labelfile=zlabelfile
        \ifeof\labelfile
        \else
            \stillreadingtrue
            \loop
                \read\labelfile to \nextline
                \ifeof\labelfile
                    \stillreadingfalse
                \else
                    \nextline
                \fi
                \ifstillreading
            \repeat
        \fi
        \immediate\openout\labelfileout=zlabelfile
    \fi}

% BE SURE TO {ENCAPSULATE} THESE CATCODE COMMANDS or the most
% amazing things will happen to your document.

{\catcode`\^^I=9
\catcode`\ =9
\catcode`\^^M=9
\endlinechar=-1
\globaldefs=1

% TAB, SPACE, and CR are to be ignored in the following macros
%    This means that when a space is needed, it is inserted by
%    the \space command.
%    All definitions are made global to get out of the encapsulation.

% This returns either a number or a letter depending on the sign
% of \chapno.  It uses the CURRENT vaule of \chapno.  Typically used
% only by other macros.  The user may however, want to place the chapter
% number (or letter) into the headline, then \chapfolio should be used.
%
\def\chapfolio{			% A <--> -2,  B <--> -3 etc.
    \ifnum \chapno>0 \relax
        \the\chapno
    \else
        \the\chapletter
    \fi}

% This increments \chapno in the correct direction (more positive OR
% more negative).  If as is normal, there is NO continuous numbering
% of equations and figures, those variables are reset.  It is typically
% used only by the other macros, not directly by the user.
%
\def\bumpchapno{
    \ifnum \chapno>-1 \relax
        \global \advance \chapno by 1
    \else
        \global \advance \chapno by -1 \setletter\chapno
    \fi
    \ifcontinuousnumbers
    \else
        \global\eqlabelno=0
    \fi
    \ifcontinuousfigurenumbers
    \else
        \global\figureno=0
    \fi
    \ifcontinuousreferencenumbers
    \else
        \global\referenceno=0
    \fi
    \sectno=0}

\def\bumpsectno{
    \global\advance\sectno by 1 \relax
    \ifeqsectionnumbers
        \ifcontinuoussectionnumbers
        \else
            \global\eqlabelno=0
        \fi
    \fi}

% This is a very awkward way to turn a number into a letter, but it's
% the only way I found that works.  [The difficulty with simpler methods
% occurred in the \write routines.]  Typically used only by other macros.
%
\def\setletter#1{\ifcase-#1 {}  \or\global\chapletter={A}
  \or\global\chapletter={B} \or\global\chapletter={C} \or\global\chapletter={D}
  \or\global\chapletter={E} \or\global\chapletter={F} \or\global\chapletter={G}
  \or\global\chapletter={H} \or\global\chapletter={I} \or\global\chapletter={J}
  \or\global\chapletter={K} \or\global\chapletter={L} \or\global\chapletter={M}
  \or\global\chapletter={N} \or\global\chapletter={O} \or\global\chapletter={P}
  \or\global\chapletter={Q} \or\global\chapletter={R} \or\global\chapletter={S}
  \or\global\chapletter={T} \or\global\chapletter={U} \or\global\chapletter={V}
  \or\global\chapletter={W} \or\global\chapletter={X} \or\global\chapletter={Y}
  \or\global\chapletter={Z}\fi}

% And a non-global version of the above:
%
\def\tempsetletter#1{\ifcase-#1 {}\or{} \or\chapletter={A} \or\chapletter={B}
 \or\chapletter={C} \or\chapletter={D} \or\chapletter={E}
  \or\chapletter={F} \or\chapletter={G} \or\chapletter={H}
   \or\chapletter={I} \or\chapletter={J} \or\chapletter={K}
    \or\chapletter={L} \or\chapletter={M} \or\chapletter={N}
     \or\chapletter={O} \or\chapletter={P} \or\chapletter={Q}
      \or\chapletter={R} \or\chapletter={S} \or\chapletter={T}
       \or\chapletter={U} \or\chapletter={V} \or\chapletter={W}
        \or\chapletter={X} \or\chapletter={Y} \or\chapletter={Z}\fi}

% A utility: it produces a number or a letter, depending on
% the sign of the argument.  Used by other macros for appendices.
% (It is like \chapfolio, but need not refer to the current chapter.)
%
\def\chapshow#1{
    \ifnum #1>0 \relax
        #1
    \else
        {\tempsetletter{\number#1}\the\chapletter}
    \fi}

% In proofmode, it is useful to put today's date and time on each output page.
%
\def\today{\number\day\space \ifcase\month\or Jan\or Feb\or
        Mar\or Apr\or May\or Jun\or Jul\or Aug\or Sep\or
        Oct\or Nov\or Dec\fi, \space\number\year}

\def\timeofday{\minutes=\time    \hours=\time
        \divide \hours by 60
        \multiply \hours by 60
        \advance \minutes by -\hours
        \divide \hours by 60
        \ifnum\the\minutes>9 \relax
     		\the\hours:\the\minutes
 	\else
  		\the\hours:0\the\minutes
	\fi}

			% either spelling

% A new chapter comes along so infrequently that it is better to handle
% the advancing of the chapter number by a separate command.  That is
%     \bumpchapno
%
%     \chapfolio		% just the number (or letter)
%     \chaplabel{...}		% set up a label for cross-reference
%
%     \chapref{...}		% in the text, refer to a labeled chapter
%
% For compatability, \chapnum is left in.  I recommend not using it.
%
\def\chapnum{\bumpchapno \chapfolio}

\def\chaplabel#1{
    \ifforwardreference                             % A label to
        \write\labelfileout{                        % reference
        \noexpand\expandafter\noexpand\def          % the chapter:
        \noexpand\csname CHAPLABEL#1\endcsname{\the\chapno}}
    \fi
    \global\expandafter\edef\csname CHAPLABEL#1\endcsname
    {\the\chapno}
    \ifproofmode
        \rlap{\hbox{\marginstyle #1\ }}
    \fi}

% For section numbers use a structure parallel to equation numbers.
% These will add to the section number and print the section number.
%
\def\sectnum{
    \bumpsectno
        \ifchaptersectionnumbers
            \chapfolio.
        \fi
    \the\sectno}

\def\sectlabel#1{
    \bumpsectno
    \ifforwardreference
        \immediate\write\labelfileout{
        \noexpand\expandafter\noexpand\def
        \noexpand\csname SECTLABEL#1\endcsname{\the\chapno.\the\sectno?!}}
    \fi
    \global\expandafter\edef\csname SECTLABEL#1\endcsname
    {\the\chapno.\the\sectno?!}	 			% automatic numbering
    \ifproofmode
        \llap{\hbox{\marginstyle #1\ }}
    \fi
    \ifchaptersectionnumbers
        \chapfolio.
    \fi
    \the\sectno}

\def\sectref#1{                                  % refer to some
    \ifundefined{SECTLABEL#1}                     % section (forward
        ++                                        % or backward):
        \ifproofmode
            \ifforwardreference
            \else
                \write16{ ***Undefined\space Section\space Reference #1*** }
            \fi
        \else
            \write16{ ***Undefined\space Section\space Reference #1*** }
        \fi
    \else
        \edef\LABxx{\getlabel{SECTLABEL#1}}
	\ifchaptersectionnumbers
            \def\LAByy{\expandafter\stripchap\LABxx}
	    \chapshow\LAByy.
	\fi
	\expandafter\stripsect\LABxx
    \fi
    \ifproofmode
        \write\allcrossfile{Section\space #1}
    \fi}

% Various ways to place a number on an equation.
%     \eqnum			% just the equation number: (7) or (3.7)
%     \eqlabel{...}		% also set up a label for reference
%     \eqalignnum  		% used in the context of \eqalignno
%     \eqalignlabel{...}	%  "    "  "     "    "      "
%
%     \eqref{...}		% in text, refer to a labeled equation
%
\def\eqnum{                                    % numbered equation
    \global\advance\eqlabelno by 1              % but no reference.
    \eqno(
    \ifeqchapternumbers
        \chapfolio.
    \fi
    \ifeqsectionnumbers
        \the\sectno.
    \fi
    \the\eqlabelno)}

\def\eqlabel#1{                                % for setting up
    \global\advance\eqlabelno by 1              % a reference to
    \ifforwardreference                     % an equation:
        \immediate\write\labelfileout{\noexpand\expandafter\noexpand\def
        \noexpand\csname EQLABEL#1\endcsname
        {\the\chapno.\the\sectno?\the\eqlabelno!}}
    \fi
    \global\expandafter\edef\csname EQLABEL#1\endcsname
    {\the\chapno.\the\sectno?\the\eqlabelno!}
    \eqno(
    \ifeqchapternumbers
        \chapfolio.
    \fi
    \ifeqsectionnumbers
        \the\sectno.
    \fi
    \the\eqlabelno)
    \ifproofmode
        \rlap{\hbox{\marginstyle #1}}		% marginal note?
    \fi}

\def\eqalignnum{                               % numbered eqalignno
    \global\advance\eqlabelno by 1              % but no reference
    &(\ifeqchapternumbers
        \chapfolio.
    \fi
    \ifeqsectionnumbers
        \the\sectno.
    \fi
    \the\eqlabelno)}

\def\eqalignlabel#1{                   	% for refering to
    \global\advance\eqlabelno by 1 	        % an EQALIGNNO'ED
    \ifforwardreference                     % equation:
        \immediate\write\labelfileout{\noexpand\expandafter\noexpand\def
        \noexpand\csname EQLABEL#1\endcsname
        {\the\chapno.\the\sectno?\the\eqlabelno!}}
    \fi
    \global\expandafter\edef\csname EQLABEL#1\endcsname
    {\the\chapno.\the\sectno?\the\eqlabelno!}
    &(\ifeqchapternumbers
        \chapfolio.
    \fi
    \ifeqsectionnumbers
        \the\sectno.
    \fi
    \the\eqlabelno)
    \ifproofmode
        \rlap{\hbox{\marginstyle #1}}			% marginal note?
    \fi}

\def\dnum{                                     % numbered equation
    \global\advance\eqlabelno by 1              % but no reference.
    \llap{(	 				% (special purpose)
    \ifeqchapternumbers
        \chapfolio.
    \fi
    \ifeqsectionnumbers
        \the\sectno.
    \fi
    \the\eqlabelno)}}

\def\dlabel#1{                                 % for setting up
    \global\advance\eqlabelno by 1              % a reference to
    \ifforwardreference                         % an equation:
        \immediate\write\labelfileout{\noexpand\expandafter\noexpand\def
        \noexpand\csname EQLABEL#1\endcsname
        {\the\chapno.\the\sectno?\the\eqlabelno!}}
    \fi
    \global\expandafter\edef\csname EQLABEL#1\endcsname
    {\the\chapno.\the\sectno?\the\eqlabelno!}
    \llap{(
    \ifeqchapternumbers
        \chapfolio.
    \fi
    \ifeqsectionnumbers
        \the\sectno.
    \fi
    \the\eqlabelno)}
    \ifproofmode
        \rlap{\hbox{\marginstyle #1}}		% marginal note?
    \fi}

\def\eqref#1{\ifparenequations(\fi
    \ifundefined{EQLABEL#1}***
        \ifproofmode
            \ifforwardreference
            \else
                \write16{ ***Undefined\space Equation\space Reference #1*** }
            \fi
        \else
            \write16{ ***Undefined\space Equation\space Reference #1*** }
        \fi
    \else
        \edef\LABxx{\getlabel{EQLABEL#1}}
	\def\LAByy{\expandafter\stripsect\LABxx}
        \def\LABzz{\expandafter\stripchap\LABxx}
        \ifeqchapternumbers
            \chapshow{\LABzz}.
        \else
            \ifnum \number\LABzz=\chapno \relax
            \else
                \chapshow{\LABzz}.
            \fi
        \fi
	\ifeqsectionnumbers
	    \LAByy.
	\fi
        \expandafter\stripeq\LABxx
    \fi
    \ifparenequations)\fi
    \ifproofmode
        \write\allcrossfile{Equation\space #1}
    \fi}

% These are analogous to \eqnum etc.  They will automatically
% generate figure numbers and in the second case accept your
% label for later reference.
%
\def\fignum{                                   % numbered figure
    \global\advance\figureno by 1\relax         % but no reference.
    \iffigurechapternumbers
        \chapfolio.
    \fi
    \the\figureno}

\def\figlabel#1{				% for setting up
    \global\advance\figureno by 1\relax 	% a reference to
    \ifforwardreference				% a figure:
        \immediate\write\labelfileout{\noexpand\expandafter\noexpand\def
        \noexpand\csname FIGLABEL#1\endcsname
        {\the\chapno.\the\sectno?\the\figureno!}}
    \fi
    \global\expandafter\edef\csname FIGLABEL#1\endcsname
    {\the\chapno.\the\sectno?\the\figureno!}
    \iffigurechapternumbers
        \chapfolio.
    \fi
    \ifproofmode
        \llap{\hbox{\marginstyle #1\ }}\relax
    \fi
    \the\figureno}

\def\figref#1{					% refer to some
    \ifundefined				% figure (forward
        {FIGLABEL#1}!!!!			% or backward):
        \ifproofmode
            \ifforwardreference
            \else
                \write16{ ***Undefined\space Figure\space Reference #1*** }
            \fi
        \else
            \write16{ ***Undefined\space Figure\space Reference #1*** }
        \fi
    \else
        \edef\LABxx{\getlabel{FIGLABEL#1}}
        \def\LABzz{\expandafter\stripchap\LABxx}
        \iffigurechapternumbers
            \chapshow{\LABzz}.\expandafter\stripeq\LABxx
        \else \ifnum\number\LABzz=\chapno \relax
                \expandafter\stripeq\LABxx
            \else
                \chapshow{\LABzz}.\expandafter\stripeq\LABxx
            \fi
        \fi
        \ifproofmode
            \write\allcrossfile{Figure\space #1}
        \fi
    \fi}

% You can refer to a specific page of the text by this macro.
% Simple say \pagelabel{yourlabel}  and you can then refer
% to that page by the name yourlabel.
%
% Note: This label making macro should
%       NOT use an \immediate\write.
%
\def\pagelabel#1{
    \ifforwardreference
        \write\labelfileout{
        \noexpand\expandafter\noexpand\def
        \noexpand\csname PGLABEL#1\noexpand\endcsname{\the\pageno}}
    \fi
    \global\expandafter\edef\csname PGLABEL#1\endcsname{\the\pageno}}

\def\pageref#1{
    \ifundefined
        {PGLABEL#1}***
        \ifproofmode
        \else
            \write16{ ***Undefined\space Page\space Reference #1*** }
        \fi
    \else
        \csname PGLABEL#1\endcsname
    \fi
    \ifproofmode
        \write\allcrossfile{Page\space #1}
    \fi}

% Another similar set of macros for footnotes:
% These are analogous to \eqnum etc.  They will automatically
% generate reference numbers and in the second case accept your
% label for later reference.
%
\def\refnum{                                      % numbered reference
    \global\advance\referenceno by 1\relax         % but no reference.
    \the\referenceno}	                           % Useless, but symmetric

\def\internalreflabel#1{			% for setting up
    \global\advance\referenceno by 1\relax 	% a reference to
    \ifforwardreference				% a reference:
        \immediate\write\labelfileout{\noexpand\expandafter\noexpand\def
        \noexpand\csname REFLABEL#1\endcsname
        {\the\chapno.\the\sectno?\the\referenceno!}}
    \fi
    \global\expandafter\edef\csname REFLABEL#1\endcsname
    {\the\chapno.\the\sectno?\the\figureno!}
    \ifproofmode
        \llap{\hbox{\marginstyle #1\hskip.5cm}}\relax
    \fi
    \the\referenceno}

\def\internalrefref#1{				% refer to some
    \ifundefined				% figure (forward
        {REFLABEL#1}!!!!			% or backward):
        \ifproofmode
            \ifforwardreference
            \else
                \write16{ ***Undefined\space Footnote\space Reference #1*** }
            \fi
        \else
            \write16{ ***Undefined\space Footnote\space Reference #1*** }
        \fi
    \else
        \edef\LABxx{\getlabel{REFLABEL#1}}
        \def\LABzz{\expandafter\stripchap\LABxx}
        \expandafter\stripeq\LABxx
        \ifproofmode
            \write\allcrossfile{Reference\space #1}
        \fi
    \fi}

% When the references are placed at the end, a standard format is used: e.g.
% These are three examples of how to make your own:
%
%\def\reflabel#1{\item{[\internalreflabel{#1}]}}
%\def\reflabel#1{\noindent{\bf \internalreflabel{#1}}.\space}
\def\reflabel#1{\item{\internalreflabel{#1}.}}

% Also, you may want a special format for placing the reference in the body
% of the text. This will make it easier to write your own.
%
%\def\refref#1{[\internalrefref{#1}]}
\def\refref#1{\internalrefref{#1}}

\def\eq{\ifhmode Eq.~\else Equation~\fi}		% most common forms.
\def\eqs{\ifhmode Eqs.~\else Equations~\fi}

% The following are typical forms for the beginning of sections:
%
% The format is   \sectionhead{The title of the section}
% OR              \sectionheadlabel{The title of the section}{alabel}
%
% The choice of \vskip size and font is typical.
% An alternative format is commented out.  This presents an underlined
% header.
%
%\def\sectionhead#1{
%    \vskip13pt plus 2pt minus 1pt
%    \centerline{
%        \ifsectionnumbers
%            \sectnum\space
%        \fi
%        $\underline{\smash{\hbox{#1}}}$}
%    \nobreak\vskip1pt\nobreak}
%
%\def\sectionheadlabel#1#2{
%    \ifsectionnumbers
%    \else
%        \message{
%    Using SECTIONHEADLABEL without SECTIONNUMBERSTRUE does not make sense.
%    }
%    \fi
%    \vskip13pt plus 2pt minus 1pt
%    \centerline{
%        \ifsectionnumbers
%            \sectlabel{#2}\space
%        \fi
%        $\underline{\smash{\hbox{#1}}}$}
%    \nobreak\vskip1pt\nobreak}

% Utilities for use by other macros
%
\def\getlabel#1{\csname#1\endcsname}
\def\ifundefined#1{\expandafter\ifx\csname#1\endcsname\relax}
\def\stripchap#1.#2?#3!{#1}			% NOTE FORMAT:
\def\stripsect#1.#2?#3!{#2}			%
\def\stripeq#1.#2?#3!{#3}			% ##.##?##!
						% chapter.section?equation!
}  %%%%% This is the termination of the special \catcode's.

\overfullrule = 0pt
\magnification = \magstep1
\baselineskip 14pt
\hsize = 6.0 truein
\vsize = 8.5 truein
%\proofmodetrue
\chapternumberstrue
%\continuousnumberingfalse
\forwardreferencetrue
\initialeqmacro
\def\sh{\mathop{\rm sh}\nolimits}

\def\tr{\mathop{\rm tr}\nolimits}

\line{\hfill UMTG-168}

%\vskip 0.2 in

\vglue 0.4 truein

\centerline{\bf MODELS WITH QUANTUM SYMMETRY AND THEIR SPECTRA
\footnote*{To appear in Louis Witten Festrschrift}}
\bigskip

\medskip

\centerline{LUCA MEZINCESCU and RAFAEL I. NEPOMECHIE}
\centerline{Department of Physics}
\centerline{University of Miami, Coral Gables, FL 33124, USA}

\vskip 0.2 in

\bigskip

\centerline{\bf ABSTRACT}

\vskip 0.2 in

{\leftskip .5truein \rightskip .5truein  \baselineskip 12pt
We review how to construct a large class of integrable quantum spin
chains with quantum-algebra symmetry, and how to determine their spectra.
\par}

\vskip 0.2 in

\bigskip

\medskip

\noindent
{\bf \chapnum . Introduction}
\vskip 0.2truein

Symmetry is one of the most reliable guiding principles of theoretical physics.
Gauge symmetries, as developed successively by Maxwell, Einstein,
Yang and Mills, and others, are just some of the many significant examples.
A new type of symmetry, called {\it quantum} symmetry, has recently
emerged${}^{\refref{qsymmetry}, \refref{faddeev}}$.
Roughly speaking, quantum Lie algebras can be regarded as one-parameter ($q$)
deformations of ordinary Lie algebras, for which one can define tensor products
of representations.

An obvious question is whether there exist interesting physical models with
quantum-algebra symmetry. A good place to look is among
integrable quantum systems${}^{\refref{qism} - \refref{kulish/sklyanin(2)}}$.
Not only were quantum algebras
first discovered in investigations of such systems, but also, the
fact that the systems are integrable has the
advantage that one may be able to determine their physical properties by
analytical (as opposed to numerical) means.

Indeed, a well-known example is
the open quantum spin chain consisting of $N$ spins with Hamiltonian
$$\eqalignno{
H &= \sum_{k=1}^{N-1} \Bigl\{ \sigma^1_k \sigma^1_{k+1} +
\sigma^2_k \sigma^2_{k+1} + {1\over 2}( q + q^{-1}) \sigma^3_k \sigma^3_{k+1}
\Bigr\} \cr
&\quad\quad\quad\quad\quad - {1\over 2}( q - q^{-1})
\Bigl( \sigma^3_1 - \sigma^3_N \Bigr) \,, \eqalignlabel{xxz} \cr}
$$
where $\vec\sigma$ are the usual Pauli matrices, and $q$ is an arbitrary
complex parameter. This model is${}^{\refref{alcaraz}, \refref{sklyanin}}$
integrable. Moreover, $H$
commutes${}^{\refref{pasquier},\refref{kulish/sklyanin(1)}}$ with the
generators $S^3$ and $S^\pm$ of the quantum algebra $U_q[su(2)]$,
$$ \left[ S^+ \,, S^-\right] = {q^{2S^3} - q^{-2S^3}\over q - q^{-1}}
\,, \quad\quad\quad \left[ S^3 \,, S^\pm \right] = \pm S^\pm \,,
\eqlabel{algebra}  $$
where
$$S^3 = \sum_{k=1}^N S^3_k \,, \quad\quad
S^\pm = \sum_{k=1}^N q^{(S^3_N + \cdots + S^3_{k+1})}\ S^\pm_k \
q^{-(S^3_{k-1} + \cdots + S^3_1)} \,, \eqlabel{comult}  $$
and
$$S^3_k = {1\over 2}\sigma^3_k \,, \quad\quad\quad
S^\pm_k = {1\over 2}(\sigma^1_k \pm i \sigma^2_k) \,. \eqlabel{spins(1)} $$
For $|q| = 1$, this model is critical and is related${}^{\refref{alcaraz},
\refref{pasquier}}$ to the $c < 1$ minimal models.

A pertinent question is whether these results can be generalized to spins
in higher-dimensional representations and to larger symmetry algebras. The
key is to formulate${}^{\refref{sklyanin},\refref{jpa}}$ open spin chains
in terms of so-called $R$ and $K$ matrices which are associated with affine
algebras $g^{(k)}$, where $g$ is a simple Lie algebra ($A_n$ = $su(n+1)$,
$B_n$ = $o(2n+1)$, $C_n$ = $sp(2n)$, $D_n$ = $o(2n)$, etc.) and
$k$ ($=1,2,3$) is the order of a diagram automorphism of $g$.

Following this approach, we have constructed${}^{\refref{ijmpa}-
\refref{miami}}$ open quantum spin chains associated with
$A^{(1)}_n$, $A^{(2)}_{2n}$, $A^{(2)}_{2n - 1}$,
$B^{(1)}_{n}$, $C^{(1)}_{n}$ and $D^{(1)}_{n}$
in the fundamental representation. These chains are integrable, and have
the quantum-algebra invariance $U_q[g_0]$, where $g_0$ is the maximal
finite-dimensional subalgebra of $g^{(k)}$. The simplest case of $A^{(1)}_1$
corresponds to the model \eqref{xxz}. (The corresponding chains with spins
in higher-dimensional representations can be constructed using a fusion
procedure${}^{\refref{kulish/sklyanin(2)},
\refref{fusion}-\refref{npb}}$.) A subset of these models, in
a different formulation, has been discussed independently in
Ref. \refref{batchelor}.

We have also shown${}^{\refref{analytical}}$ how the spectra of these chains
(with the exception of $A^{(1)}_n$ for $n>1$)
can be exactly determined -- albeit implicitly through solutions of Bethe
Ansatz equations -- using a generalization of the analytical Bethe Ansatz
method. In this approach${}^{\refref{reshetikhin}}$, one uses general
properties of the $R$ matrix (such as analyticity, unitarity, crossing
symmetry, etc.) to derive various properties of the transfer-matrix
eigenvalues.
These properties are used to completely determine the eigenvalues,
assuming that they have the form of ``dressed'' pseudovacuum eigenvalues.

Here we summarize these results. Specifically, in Sec. 2 we present the
general model, and state its two important properties:
integrability and quantum-algebra invariance. In Sec. 3, we explain how the
spectrum can be determined. Sec. 4 contains some concluding remarks.

\vskip 0.4truein

\noindent
{\bf \chapnum .  The model}

\vskip 0.2truein

We consider the {\it open} quantum spin chain with Hamiltonian
$$H = \sum_{k=1}^{N-1} {d\over du} \check R_{k,k+1}(u) \Big\vert_{u=0}
\,, \eqlabel{hamiltonian} $$
whose state space is $\otimes^N \ { V}$, where ${ V}$ is some
finite-dimensional linear vector space.
The notations here are standard. (See, e.g., Refs. \refref{kulish/sklyanin(3)}
and \refref{kulish/sklyanin(2)}.)
In particular, $\check R(u) = {\cal P} R(u)$, where
${\cal P}$ is the permutation matrix in ${ V} \otimes { V}$
(i.e., ${\cal P} (x \otimes y) = y \otimes x$ for $x, y \in { V}$),
and $R(u)$ is a matrix acting in ${ V} \otimes { V}$
which obeys the Yang-Baxter equation
$$
R_{12}(u - v)\  R_{13}(u)\ R_{23}(v) = R_{23}(v)\  R_{13}(u)\ R_{12}(u - v)
\,.   \eqlabel{yang-baxter}
$$
As usual, $R_{12}(u)$, $R_{13}(u)$ and $R_{23}(u)$ are matrices acting
in ${ V} \otimes { V} \otimes { V}$, with $R_{12}(u) = R(u)
\otimes 1$, $R_{23}(u) = 1 \otimes R(u)$, etc.

We require that $R$ have certain additional properties.
These properties are in fact satisfied by a subset of the $R$ matrices
associated with affine algebras $g^{(k)}$, as detailed in the following Lemma:

\medskip
\noindent {\it Lemma 1.} {\it The $R$ matrices
associated with $A^{(1)}_n$, $A^{(2)}_{2n}$, $A^{(2)}_{2n - 1}$,
$B^{(1)}_{n}$, $C^{(1)}_{n}$ and $D^{(1)}_{n}$
in the fundamental representation have the following properties:\hfill\break
$PT$ symmetry}
$${\cal P}_{12}\ R_{12}(u)\ {\cal P}_{12} \equiv R_{21}(u)
= R_{12}(u)^{t_1 t_2} \,,
\eqlabel{pt}
$$
{\it unitarity}
$$R_{12}(u)\  R_{21}(-u) = \zeta (u) \,, \eqlabel{unitarity} $$
{\it and regularity}
$$R_{12}(0) = \zeta(0)^{1/2}\ {\cal P}_{12} \,; \eqlabel{regularity} $$
{\it also,}
$$\{\{\{R_{12}(u)^{t_2}\}^{ -1}\}^{t_2}\}^{-1}
= {\zeta(u+\rho)\over \zeta(u + 2\rho)}
M_2\ R_{12}(u + 2\rho) M_2^{-1} \,, \eqlabel{m1} $$
$$ \left[ R(u) \,, M \otimes M \right] = 0 \,; \eqlabel{m2} $$
{\it and finally,}
$$\left[ \check R_{12}(u) \,, \check R_{12}(v) \right] = 0  \,.
\eqlabel{Rcheck}  $$

\medskip

We use here the following notations: $t_i$ denotes transposition in the
$i^{th}$ space; and $M_1 = M \otimes 1$, $M_2 = 1 \otimes M$. The quantity
$\zeta (u)$ which is introduced in \eq\eqref{unitarity}
is some even scalar function of $u$; the quantity $M$ which appears
in Eqs. \eqref{m1} and \eqref{m2} is some symmetric matrix
($M^t = M$); and $\rho$ is some constant.

The properties \eqref{pt} - \eqref{regularity} were
noted by Bazhanov${}^{\refref{bazhanov}}$ and by Jimbo${}^{\refref{jimbo}}$;
the properties \eqref{m1} and \eqref{m2}
were noted by Reshetikhin and Semenov-Tian-Shansky${}^{\refref{semenov}}$;
and the property \eqref{Rcheck} was noted by Jimbo${}^{\refref{jimbo}}$.
This last property is gauge-dependent, and is valid in the so-called
homogeneous gauge.
While there are affine algebras (such as $D^{(2)}_{n}$ ) for which the
associated $R$ matrices are known not to have the last property \eqref{Rcheck},
the other properties (\eqref{pt} - \eqref{m2}) are presumably of more general
validity. We emphasize that we do {\it not} require that $R$ have crossing
symmetry.

We define the so-called transfer matrix $t(u)$ as follows:
$$t(u) = \tr_a M_a\ T_a(u)\  \hat T_a(u) \,, \eqlabel{transfer} $$
where $M$ is the matrix that was introduced above, and
$$\eqalignno{
     T_a(u) &= R_{aN}(u)\ R_{a N-1}(u)\ \cdots R_{a1}(u) \,, \cr
\hat T_a(u) &= R_{1a}(u)\ \cdots R_{N-1 a}(u)\ R_{Na}(u) \,.
\eqalignlabel{monodromy} \cr} $$
The subscript $a$ denotes the auxiliary space. As usual, we suppress the
quantum-space subscripts $1 \,, \cdots \,, N$ of $T_a(u)$ and $\hat T_a(u)$.

\medskip
\noindent
{\it Lemma 2.} {\it For the $R$ matrices listed in Lemma 1, the
transfer matrix $t(u)$ constitutes a one-parameter commutative family}
$$\left[ t(u) \,, t(v) \right] = 0 \hbox{   for all   } u \,, v
\,. \eqlabel{commutativity} $$
{\it This transfer matrix is related to the Hamiltonian \eqref{hamiltonian} by}
$$H = \gamma {d\over du} t(u) \Big\vert_{u=0} \,, \quad\quad\quad
\gamma = {1\over 2 \zeta(0)^{N - {1\over 2}}\left( \tr_a M_a \right)}
\,, \eqnum $$
{\it and the model defined by this Hamiltonian is integrable.}

\medskip

We prove this Lemma in Ref. \refref{ijmpa}, following the approach developed
by Sklyanin in Ref. \refref{sklyanin} for open integrable chains. The
integrability of such chains originates from a larger algebraic structure
which is described in these papers.

\noindent
{\it Remark.} An equivalent expression for the transfer matrix
is${}^{\refref{analytical}}$
$$t(u) = \tr_a M_a^{-1}\ \hat T_a(u)\ T_a(u)
\,. \eqlabel{transfertoo} $$
This result is important for Sec. 3.

We next turn to the quantum-algebra symmetry of the model. We
recall (see, e.g., Ref. \refref{faddeev}) that
for an $R$ matrix associated with the affine algebra $g^{(k)}$,
the leading asymptotic behavior for large $u$ is given by
$$R_{ab}(u) \sim \iota(u)\ R^+_{ab} \quad\quad {\hbox{  for  }}
u \rightarrow \infty\,,
\eqnum $$
where $\iota(u)$ is a scalar function of $u$, and
$R^+_{ab}$ is a $u$-independent triangular matrix.
It follows that
$$T_a(u) \sim \kappa(u)\ T^+_a \,, \quad\quad\quad
\hat T_a(u) \sim \kappa(u)\ \hat T^+_a \quad\quad {\hbox{  for  }}
u \rightarrow \infty\,.  \eqlabel{asymptoticT} $$
Here $\kappa(u) = \iota(u)^N$ , and
$T^+_a$ and $\hat T^+_a$ are $u$-independent
triangular matrices which satisfy the algebra
$$\eqalignno{
R^+_{ab}\ T^+_a\ T^+_b &= T^+_b\ T^+_a\ R^+_{ab} \,, \cr
\hat T^+_b\ R^+_{ab}\ T^+_a\  &= T^+_a\ R^+_{ab}\ \hat T^+_b \,, \cr
R^+_{ab}\ \hat T^+_b\ \hat T^+_a &= \hat T^+_a\ \hat T^+_b\ R^+_{ab} \,.
\eqalignnum \cr} $$
The operators $T^+_a$ and $\hat T^+_a$ can be
expressed in terms of generators of the quantum algebra $U_q[g_0]$,
where $g_0$ is the maximal finite-dimensional subalgebra of $g^{(k)}$.
The diagonal entries of $T^+_a$ and $\hat T^+_a$ involve the Cartan
generators, while the off-diagonal elements are either raising or lowering
operators.

\medskip
\noindent
{\it Lemma 3.} {\it For the $R$ matrices listed in Lemma 1,
the transfer matrix $t(u)$ commutes with the operators $T^+_a$ and
$\hat T^+_a$,}
$$\Bigl[ t(u) \,,  T^+_a  \Bigr] = 0 \,, \quad\quad\quad
  \Bigl[ t(u) \,,  \hat T^+_a  \Bigr] = 0 \,, \eqlabel{commute}  $$
{\it and therefore}
$$ \Bigl[ t(u) \,, U_q [g_0] \Bigr] = 0 \,. \eqlabel{qinvariance}  $$

\medskip

That is, not only the Hamiltonian \eqref{hamiltonian} but also the
transfer matrix \eqref{transfer} commutes with generators of a quantum
algebra. We prove this Lemma in Ref. \refref{mpla}
(see also Ref. \refref{analytical} ), following the argument of
Kulish and Sklyanin${}^{\refref{kulish/sklyanin(1)}}$ for the
$A^{(1)}_1$ case.

\medskip
\noindent
{\it Example.} As an example, we consider the $A^{(1)}_1$ case.
The $R$ matrix is given by
$$R(u)= \left( \matrix{ \sh(u+\eta)                         \cr
                         &  \sh u          & e^u \sh \eta  \cr
                         & e^{-u} \sh \eta  & \sh u        \cr
                         &                  & & \sh(u+\eta) \cr} \right) \,.
\eqlabel{R(1)} $$
This form of the $R$ matrix differs from the more familiar symmetric form
(which
does not satisfy the commutativity property \eqref{Rcheck})
by a gauge transformation${}^{\refref{jimbo}, \refref{gauge},\refref{jpa}}$.

Evaluating the Hamiltonian \eqref{hamiltonian} for this $R$ matrix, we recover
the model \eqref{xxz}, with $q=e^\eta$. Furthermore, the asymptotic behavior
of the monodromy matrices $T_{a}(u)$ and $\hat T_{a}(u)$ for large $u$ is
given by \eqref{asymptoticT}, with
$$\kappa(u) = \left({1\over 2}\right)^N  e^{(u + {\eta\over 2}) N} \,, $$
$$T^+_a  =
\left( \matrix{ e^{\eta S^3}  & p\ S^- \cr
                     0         & e^{-\eta S^3} \cr} \right) \,, \quad\quad\quad
\hat T^+_a =
\left( \matrix{ e^{\eta S^3}  & 0  \cr
p\ S^+  & e^{-\eta S^3} \cr} \right) \,,  \quad\quad\quad
p = 2 e^{-{\eta\over 2}} \sh \eta \,, \eqlabel{largeu}
$$
and $S^3$, $S^\pm$ are given by \eqref{comult}.
(In obtaining these results, we have written the $R$ matrix \eqref{R(1)}
as a $2 \times 2$ matrix in the auxiliary space, with operator entries.)

As expected, $T^+_a$ and $\hat T^+_a$ are triangular matrices whose entries
are generators of $U_q[su(2)]$. The fact that $t(u)$ commutes with
$T^+_a$ and $\hat T^+_a$ implies that $t(u)$ commutes with the
generators of $U_q[su(2)]$.

\vskip 0.4truein
\noindent
{\bf \chapnum . Analytical Bethe Ansatz}
\vskip 0.2truein

The fact that the model \eqref{hamiltonian} is integrable indicates
that it can be exactly solved. The analytical Bethe Ansatz
is an approach developed by Reshetikhin${}^{\refref{reshetikhin}}$
to determine the eigenvalues of the transfer matrix of a closed
integrable chain. In this approach, one makes use of further properties of
the $R$ matrix, including crossing symmetry
$$R_{12}(u) = V_1 \ R_{12}(-u - \rho)^{t_2}\  V_1
            = V_2^{t_2} \ R_{12}(-u - \rho)^{t_1}\  V_2^{t_2}  \,,
\eqlabel{crossing} $$
where $V$ is a matrix which satisfies $V^2 = 1$. This property is stronger
than \eqref{m1}, and in fact implies that
$$M = V^t\ V  \,. \eqnum  $$
All the $R$ matrices listed in Lemma 1 have crossing symmetry,
with the exception of $A^{(1)}_n$ for $n>1$. A generalization of the analytical
Bethe Ansatz for {\it open} integrable chains can presumably be applied to
the remaining cases. We work out in detail the $A^{(1)}_1$ and $A^{(2)}_2$
cases in Ref. \refref{analytical}.

In order to illustrate the method, we review here the $A^{(1)}_1$ case.
For real values of $\eta$, the $R$ matrix \eqref{R(1)} is real, and hence
$t(u)$
is Hermitian. Since $t(u)$ commutes with the generators of $U_q[su(2)]$,
there exist simultaneous eigenstates $|\Lambda^{(m)}>$ of $t(u)$ and ${\cal
M}$,
$$\eqalignno{
    t(u)\ |\Lambda^{(m)}> &= \Lambda^{(m)}(u)\ |\Lambda^{(m)}> \,, \cr
{\cal M}\ |\Lambda^{(m)}> &= m\ |\Lambda^{(m)}> \,, \eqalignnum \cr} $$
where
$$ {\cal M} = {N\over 2} - S^3 \,. \eqlabel{calM(1)} $$
We choose the eigenstates to be highest weights of $U_q[su(2)]$,
$$ S^+\ |\Lambda^{(m)}> = 0 \,. \eqlabel{highestweight(1)} $$
(Within the algebraic Bethe Ansatz approach, one can
prove${}^{\refref{mpla}, \refref{devega}}$ that the Bethe states of this
model are highest weights of $U_q[su(2)]$.)

Our task is to determine the eigenvalues $\Lambda^{(m)}(u)$.
To this end, we first consider the so-called pseudovacuum state which has
all $N$ spins up:
$$|\uparrow \uparrow \cdots \uparrow> = \prod_{k=1}^N \otimes |\uparrow>_k \,,
\eqlabel{pseudovacuum} $$
where
$$|\uparrow>_k = \left( \matrix{1 \cr
                                0 \cr} \right)_k \,. \eqnum $$
This state clearly satisfies the highest-weight
condition \eqref{highestweight(1)}, and has $m=0$. Let us assume that
this is a normalized eigenstate $|\Lambda^{(0)}>$ of $t(u)$.
We compute the expectation value of $t(u)$ with respect to this state,
using the fact that $<\Lambda^{(0)}|\ T_{a}(u)$ and
$\hat T_{a}(u)\ |\Lambda^{(0)}>$ are lower- and upper- triangular matrices,
respectively. In this way, we obtain the corresponding eigenvalue
$\Lambda^{(0)}(u)$,
$$\Lambda^{(0)}(u) = -{1\over \sh(2u + \eta)}
\left\{ \sh(2u + 2\eta)\ \sh^{2N}(u + \eta)
+ \sh2u\ \sh^{2N} u \right\} \,. \eqnum $$

Assuming that a general eigenvalue $\Lambda^{(m)}(u)$ has the form of a
``dressed'' pseudovacuum eigenvalue, we are led to the following analytical
Ansatz,
$$\eqalignno{
\Lambda^{(m)}(u) =-{1\over \sh(2u + \eta)} \Bigl\{
& A^{(m)}(u)\ \sh(2u + 2\eta)\ \sh^{2N}(u + \eta) \cr
& \ + B^{(m)}(u)\ \sh2u\ \sh^{2N}u \Bigr\} \,.
\eqalignlabel{ansatz(1)} \cr} $$
The problem now is to determine the unknown functions $A^{(m)}(u)$ and
$B^{(m)}(u)$ .

With the help of \eqref{largeu}, we find that the leading
asymptotic behavior of $\Lambda^{(m)}(u)$ for large $u$ is given by
$$\Lambda^{(m)}(u) \sim - \left({1\over 2}\right)^{2N}  e^{2 u N}
\left\{ e^{\eta (1 + 2N - 2m)} + e^{\eta (-1 + 2m)}  \right\} \,.
\eqlabel{asymptotic(1)} $$
It follows that for $u \rightarrow \infty$,
$$A^{(m)}(u) \rightarrow e^{-2m\eta} \,, \quad\quad\quad
  B^{(m)}(u) \rightarrow e^{2m\eta}  \,. \eqlabel{ABasymptotic} $$

The $R$ matrix \eqref{R(1)} has the property
$$R(u + i\pi) = - R(u) \,, \eqnum $$
which implies that the transfer matrix is a periodic function
of $u$, with period $i\pi$,
$$t(u + i\pi) = t(u) \,. \eqlabel{periodicity(1)} $$
Therefore $A^{(m)}(u)$ and $B^{(m)}(u)$ have the same periodicity
property. This fact will help us to determine the $u$-dependence
of these functions.

In order to proceed further, we make use of two functional
equations for the eigenvalues. The first equation expresses
the fusion property of the transfer matrix. Indeed, by fusing in the
auxiliary space, we construct a ``fused'' transfer matrix
$\tilde t(u)$ which is related to the original transfer matrix $t(u)$ by the
fusion formula${}^{\refref{npb}}$
$$\tilde t(u) = a(u) \ t(u)\ t(u+\rho) + b(u) \,. \eqlabel{fusion-transfer} $$
Here $a(u)$ and $b(u)$ are known scalar functions, which for $A^{(1)}_1$
are given by
$$\eqalignno{
a(u) &= - {\sh (2u + 3\eta)\ \sh(2u + \eta) \over \sh^{2N} (u + \eta)\
\sh^2 (2u + 2\eta)} \,, \cr
b(u) &= {\sh 2u\ \sh(2u + 4\eta) \over \sh^2 (2u + 2\eta)}
\left[ {\sh u\ \sh(u + 2\eta) \over \sh (u + \eta)} \right]^{2N}
\,, \eqalignnum \cr}$$
with $\rho = \eta + i\pi$.
Acting with the fusion formula on an eigenstate $|\Lambda^{(m)}>$ of $t(u)$,
we obtain
$$\tilde \Lambda^{(m)}(u) = a(u)\ \Lambda^{(m)}(u)\ \Lambda^{(m)}(u+\rho)
+ b(u) \,, \eqlabel{fusion-lambda} $$
which, as we shall see, can be regarded as a functional equation for
$\Lambda^{(m)}(u)$.

The transfer matrix obeys the crossing relation
$$t(u) = t(-u -\rho) \,.  \eqlabel{crossing-transfer} $$
This relation, which we prove with the help of crossing symmetry
\eqref{crossing} and the second form \eqref{transfertoo} of the
transfer matrix, leads to a second functional equation
$$\Lambda^{(m)}(u) = \Lambda^{(m)}(-u -\rho) \,. \eqlabel{crossing-lambda} $$

We substitute the above Ansatz into the fusion equation
\eqref{fusion-lambda}, and we demand the cancelation of the $b(u)$
term. The fact that this cancelation occurs for
$A^{(0)} = B^{(0)} = 1$ confirms our earlier assumption that the
pseudovacuum state is an eigenstate of $t(u)$. Moreover, for general
values of $m$, we obtain the result
$$A^{(m)}(u + \rho)\ B^{(m)}(u) = 1 \,. \eqlabel{ABfusion} $$
Similarly,
substituting the Ansatz into the crossing relation
\eqref{crossing-lambda}, we obtain
$$ A^{(m)}(u) = B^{(m)}(-u - \rho) \,. \eqlabel{ABcrossing} $$
Combining these two results, we deduce that
$$A^{(m)}(u)\ A^{(m)}(-u) = 1 \,. \eqlabel{AA(1)} $$

One can argue that $\Lambda^{(m)}(u)$ is a finite power series in
$e^{2u}$; in particular, it has poles only at
$e^{2u} = 0$ and at $e^{2u} = \infty$.
In view of the Ansatz \eqref{ansatz(1)}, we see that $A^{(m)}$ and $B^{(m)}$
must be rational functions of $e^{2u}$; they must have common poles, and the
residues of $\Lambda^{(m)}(u)$ at these poles must vanish.
Imposing also the asymptotic behavior \eqref{ABasymptotic} and the relations
\eqref{ABcrossing} and \eqref{AA(1)},
we eventually conclude that the eigenvalues $\Lambda^{(m)}(u)$ are given by
\eqref{ansatz(1)}, with
$$\eqalignno{
A^{(m)}(u) & = \prod_{j=1}^m
{\sh(u - u_j - {\eta\over 2}) \over \sh(u - u_j + {\eta\over 2})}
{\sh(u + u_j - {\eta\over 2}) \over \sh(u + u_j + {\eta\over 2})}  \,, \cr
B^{(m)}(u) & = \prod_{j=1}^m
{\sh(u - u_j + {3\eta\over 2}) \over \sh(u - u_j + {\eta\over 2})}
{\sh(u + u_j + {3\eta\over 2}) \over \sh(u + u_j + {\eta\over 2})} \,,
\eqalignlabel{spectrum(1)} \cr} $$
where $\left\{u_1 \,, \cdots u_m \right\}$ are solutions of the
Bethe Ansatz (BA) equations
$$\left[ {\sh(u_k + {\eta\over 2}) \over \sh(u_k - {\eta\over 2})} \right]^{2N}
= \prod_{j \ne k} {\sh(u_k - u_j + \eta) \over \sh(u_k - u_j - \eta)}
{\sh(u_k + u_j + \eta) \over \sh(u_k + u_j - \eta)} \,, \quad\quad
k = 1\,, \cdots \,, m \,. \eqlabel{BA(1)} $$
These
results coincide with those obtained in
Refs. \refref{alcaraz} and \refref{sklyanin} using the coordinate BA and
the algebraic BA, respectively. We have worked out the $A^{(2)}_2$ case
in a similar manner.

\vskip 0.4truein
\noindent
{\bf \chapnum . Discussion}
\vskip 0.2truein

Comparing the BA equations \eqref{BA(1)} with those of the
corresponding closed chain with periodic boundary conditions
(which is {\it not} $U_q[su(2)]$-invariant),
we see that the former are ``doubled'' with respect to the latter.
Moreover, the transfer-matrix eigenvalues of the
quantum-algebra-invariant chain are also doubled with respect to the
eigenvalues of the corresponding closed chain, up to certain prefactors.

Evidently, this doubling phenomenon is related to the fact that the
open-chain transfer matrix \eqref{transfer} involves both $T_a(u)$ and
$\hat T_a(u)$, while the closed-chain transfer matrix
involves only $T_a(u)$.
Within the analytical BA approach, the signal of this doubling phenomenon
is the existence of a crossing relation \eqref{crossing-transfer} for the
transfer matrix. Since such a
relation holds for all the cases enumerated in Lemma 1 except
$A^{(1)}_n$ ($n > 1$), we expect that the doubling phenomenon should
occur for at least these cases.

A number of interesting questions remain:
\item{1.} How to construct integrable quantum-algebra-invariant
chains associated with {\it twisted} affine $D_n$ algebras, for which
the commutativity property \eqref{Rcheck} does not hold?
Moreover, the exceptional cases have not yet been investigated.
\item{2.} How to solve the $A^{(1)}_n$ ($n > 1$) chains? Presumably,
one can generalize the so-called nested${}^{\refref{nested}}$ BA approach.
\item{3.} What are the thermodynamic ($N \rightarrow \infty$) properties of
these models? This is most interesting and subtle in the critical regime
$|q|=1$, since in this case the Hamiltonian is typically not Hermitian, and
presumably truncations on the space of states are required.

\vskip 0.4truein
\noindent
{\bf \chapnum . Acknowledgements}
\vskip 0.2truein

%We dedicate this paper to Professor Louis Witten.
We thank the Editors for inviting us to the Symposium,
and for giving us the honor to contribute to this volume.
This work was supported in part by the National Science
Foundation under Grant No. PHY-90 07517.

\vskip 0.4truein
\noindent
{\bf \chapnum . References}
\vskip 0.2truein

\reflabel{qsymmetry}
V.G. Drinfel'd, J. Sov. Math. {\it 41} (1988) 898;
M. Jimbo, Int'l J. Mod. Phys. {\it A4} (1989) 3759.

\reflabel{faddeev}
L.D. Faddeev, N. Yu. Reshetikhin and L.A. Takhtajan, Algebraic Analysis,
{\it 1} (1988) 129;  Leningrad Math. J. {\it 1} (1990) 193.

\reflabel{qism}
R.J. Baxter, {\it Exactly Solved Models in Statistical Mechanics}
(Academic Press, 1982);
L.D. Faddeev and L.A. Takhtajan, Russ. Math. Surv. {\it 34} (1979) 11;
J. Sov. Math. {\it 24} (1984) 241;
L.D. Faddeev, in {\it Les Houches Lectures 1982} ed. by J.B. Zuber and R. Stora
(North-Holland, 1984) 561;
H.J. de Vega, Int'l J. Mod. Phys. {\it A4} (1989) 2371.

\reflabel{kulish/sklyanin(3)}
P.P. Kulish and E.K. Sklyanin, J. Sov. Math. {\it 19} (1982) 1596.

\reflabel{kulish/sklyanin(2)}
P.P. Kulish and E.K. Sklyanin, {\it Lecture Notes in Physics} {\it 151}
(Springer, 1982) 61.

\reflabel{alcaraz}
F.C. Alcaraz, M.N. Barber, M.T. Batchelor, R.J. Baxter and G.R.W. Quispel,
J. Phys. {\it A20} (1987) 6397.
See also M. Gaudin, Phys. Rev. {\it A4} (1971) 386;
{\it La fonction d'onde de Bethe} (Masson, 1983).

\reflabel{sklyanin}
E.K. Sklyanin, J. Phys. {\it A21} (1988) 2375.
See also A.B. Zamolodchikov, unpublished;
I.V. Cherednik, Theor. Math. Phys. {\it 61} (1984) 977.

\reflabel{pasquier}
V. Pasquier and H. Saleur, Nucl. Phys. {\it B330} (1990) 523.

\reflabel{kulish/sklyanin(1)}
P.P. Kulish and E.K. Sklyanin, J. Phys. {\it A24} (1991) L435 ; in
{\it Proc. Euler Int. Math. Inst., 1st Semester: Quantum Groups, Autumn 1990},
ed. by P.P. Kulish, in press.

\reflabel{jpa}
L. Mezincescu and R.I. Nepomechie, J. Phys. {\it A24} (1991) L17.

\reflabel{ijmpa}
L. Mezincescu and R.I. Nepomechie, Int'l J. Mod. Phys. {\it A6} (1991) 5231;
Addendum, in press.

\reflabel{mpla}
L. Mezincescu and R.I. Nepomechie, Mod. Phys. Lett. {\it A6} (1991) 2497.

\reflabel{miami}
L. Mezincescu and R.I. Nepomechie, in {\it Quantum Field Theory,
Statistical Mechanics, Topology and Quantum Groups}, ed. by T.L. Curtright,
L. Mezincescu and R.I. Nepomechie, in press.

\reflabel{fusion}
M. Karowski, Nucl. Phys. {\it B153} (1979) 244;
P.P. Kulish, N.Yu. Reshetikhin and E.K. Sklyanin, Lett. Math. Phys. {\it 5}
(1981) 393.

\reflabel{spin1}
L. Mezincescu, R.I. Nepomechie and V. Rittenberg, Phys. Lett. {\it A147}
(1990) 70;
L. Mezincescu and R.I. Nepomechie, in {\it Argonne Workshop
on Quantum Groups}, ed. by T.L. Curtright, D. Fairlie and C. Zachos
(World Scientific, 1991) 206.

\reflabel{npb}
L. Mezincescu and R.I. Nepomechie, J. Phys. {\it A25} (1992) 2533.

\reflabel{batchelor}
M.T. Batchelor and A. Kuniba, J. Phys. {\it A24} (1991) 2599.

\reflabel{analytical}
L. Mezincescu and R.I. Nepomechie, Nucl. Phys. {\it B372} (1992) 597.

\reflabel{reshetikhin}
V.I. Vichirko and N.Yu. Reshetikhin, Theor. Math. Phys. {\it 56}
(1983) 805; N.Yu. Reshetikhin, Lett. Math. Phys. {\it 7} (1983) 205;
Sov. Phys. JETP {\it 57} (1983) 691; Lett. Math. Phys. {\it 14} (1987) 235.

\reflabel{bazhanov}
V.V. Bazhanov, Phys. Lett. {\it 159B} (1985) 321;
Commun. Math. Phys. {\it 113} (1987) 471.

\reflabel{jimbo}
M. Jimbo, Commun. Math. Phys. {\it 102} (1986) 537;
{\it Lecture Notes in Physics}, Vol. 246 (Springer, 1986) 335.

\reflabel{semenov}
N. Yu. Reshetikhin and M.A. Semenov-Tian-Shansky, Lett. Math. Phys.
{\it 19} (1990) 133. See also I.B. Frenkel and N. Yu. Reshetikhin,
Commun. Math. Phys. {\it 146} (1992) 1.

\reflabel{gauge}
K. Sogo, Y. Akutsu and T. Abe, Prog. Theor. Phys. {\it 70} (1983) 730.

\reflabel{devega}
C. Destri and H.J. de Vega, Nucl. Phys. {\it B}, in press.

\reflabel{nested}
C.N. Yang, Phys. Rev. Lett. {\it 19} (1967) 1315;
P.P. Kulish and N.Yu. Reshetikhin, Sov. Phys. JETP {\it 53} (1981) 108;
O. Babelon, H.J. de Vega and C-M. Viallet, Nucl. Phys. {\it B200} (1982) 266.

\end